\documentclass[twocolumn,pra,showpacs,amsmath,amssymb,superscriptaddress]{revtex4-1}

\usepackage{graphicx}
\usepackage{dcolumn}
\usepackage{amsmath}
\usepackage[hypertex]{hyperref}
\usepackage{amsfonts}
\usepackage{amsthm}
\usepackage{times}
\usepackage{amssymb}
\usepackage{mathrsfs}
\usepackage{graphicx}
\PassOptionsToPackage{caption=false}{subfig}
\usepackage{subfig}
\usepackage{color}

\begin{document}

\title{Hybrid Qubit gates in circuit QED: A scheme for quantum bit encoding and information processing}

\author{O. P. de S\'a Neto}
 \email{opsn@ifi.unicamp.br}
\affiliation{Instituto de F\'{\i}sica Gleb Wataghin, Universidade Estadual de Campinas, CEP 13083-859, Campinas, S\~{a}o Paulo,
Brazil}

\author{M. C. de Oliveira}
 \email{marcos@ifi.unicamp.br}
\affiliation{Instituto de F\'{\i}sica Gleb Wataghin, Universidade Estadual de Campinas, CEP 13083-859, Campinas, S\~{a}o Paulo,
Brazil}
\affiliation{Institute for Quantum Information Science, University of Calgary, Alberta T2N 1N4, Canada}

\date{\today}

\begin{abstract}
{Solid state superconducting devices coupled to coplanar transmission lines offer an exquisite architecture for quantum optical phenomena probing as well as for quantum computation implementation, being the object of intense theoretical and experimental investigation lately.  In appropriate conditions the transmission line radiation modes can get strongly coupled to a superconducting device with only two levels -for that reason called artificial atom or qubit. Employing this system we propose a hybrid two-quantum bit gate encoding involving quantum electromagnetic field qubit states prepared in a coplanar transmission line capacitively coupled to a single charge qubit. Since dissipative effects are more drastic in the solid state qubit than in the field one, it can be employed for storage of information, whose efficiency against the action of an ohmic bath show that this encoding can be readily implemented with present day technology. We extend the investigation to generate entanglement between several solid state qubits and the field qubit through the action of external classical magnetic pulses.}

\end{abstract}

\pacs{03.67.Mn, 03.65.Ud, 03.65.Yz, } \maketitle

\section{INTRODUCTION}
Cavity Quantum Electrodynamics (QED) traditionally deals with the interaction between quantized radiation fields and two or more atomic levels. Many features can be simply formulated by considering only two levels of the atomic system and its interaction with a quantized field inside the dipole approximation \cite{Milburn} as given by \begin{equation}
H=\hbar \omega_{c}a^{\dag}a+\hbar \nu_{a}\sigma_{z}+\hbar g\sigma_{x} \left(a^{\dag}+a\right)+H_{\gamma}+H_{\kappa},\label{eq1}
\end{equation}
where $a$($a^{\dag}$) are bosonic annihilation (creation) operators, $\nu_{a}$ is the frequency of the atomic transition, $g$ is the coupling constant between atom-field, $\sigma_{z}=\left|g\right\rangle\left\langle g\right|-\left|e\right\rangle\left\langle e\right|$, $\sigma_{x}=\left|e\right\rangle\left\langle g\right|+\left|g\right\rangle\left\langle e\right|$,  where $\left|g\right\rangle$ is the ground atomic state, $\left|e\right\rangle$ is the excited atomic state. In Eq. (\ref{eq1}) $H_{\gamma}$ is the Hamiltonian employed for describing the relaxation of the system due to the coupling of the atom to the modes of the vacuum, and $H_{\kappa}$ is  the Hamiltonian responsible for the decay of the photons in the cavity field. Important results within this model and inside the rotating wave approximation have been developed in the past  \cite{Milburn,Scully,Yamamoto}. One particularly important result was the generation of superposition states obtained  by  Haroche's group since 1997 \cite{H,ref2}, with Rydberg atoms in microwave cavities, creating superposition states of coherent state of the field.  In order to generate superposition states, changes in populations are undesirable for the operation. For this to be satisfied the regime of strong dispersive interaction must be reached, possible only in the microwave spectrum of the radiation field and involving Rydberg atoms, and more recently on circuit QED \cite{mae,mae1}. 
\begin{figure}[!ht]
\setlength\fboxsep{0.2pt}
\setlength\fboxrule{0.2pt}{\includegraphics[scale=0.45]{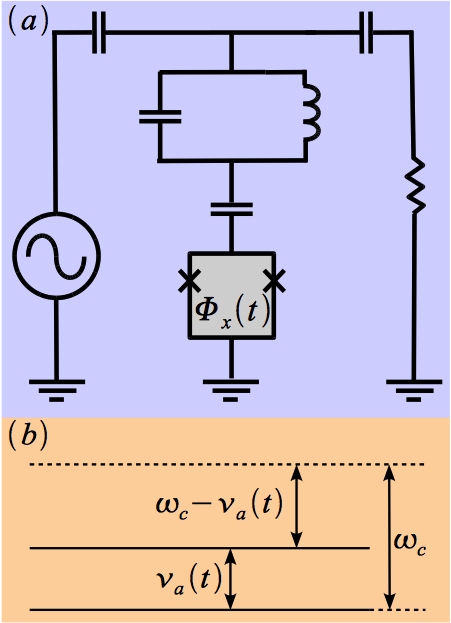}}
\caption{\label{fig1}($a$) Schematic of the proposal, with the central transmission line (resonator) capacitively coupled to the source and the drain of the quantized electromagnetic radiation, and also coupled capacitively to the solid state qubit. ($b$) Illustration of the detuning of the frequency of the quantized radiation  from the qubit transition frequency in the \textit{blue detuned} regime.}
\end{figure}

Circuit QED with superconducting devices has several advantages over cavity QED experiments involving Rydberg atoms - With superconducting circuit elements the coherence time of the field in transmission line resonators is very large; there is a strong "artificial atom"-field coupling, convenient for the exchange of information. Thus, a regime of strong dispersive interaction otherwise impossible to be attained in experiments with atoms, as well as with quantum dots, can be reached in a very convenient assembling into a chip. This feature turns this system quite attractive for quantum information processing \cite{eeq,qubit}, and many proposal have been addressed recently ( See e.g. \cite{solano1,korotkov,pashkin,omc,JG,entq,analog,solano} and references therein).
The circuit corresponding to Eq.(\ref{eq1}) is schematized in Fig.    \ref{fig1}, where a central resonator transmission line is capacitively coupled to a charge qubit. The resonator is initially fed by an AC field, so that a coherent field is stablished in a fundamental mode, and the charge qubit transition frequency is adjusted by a time dependent external classical magnetic flux. Usually the field is the carrier for quantum information encoded in solid state qubits, but it is possible to employ the field itself in the encoding of qubits. For that one has to envision a good encoding in terms of a robust set of mutually orthogonal field states. Unfortunately, single and zero photon (Fock) states are too sensitive to dissipative effects \cite{Fock,Fock2}. A better alternative is to employ a robust set of orthogonal superpositions of coherent states, known as odd and even coherent states \cite{Mc1}. Those states have been historically proved hard to generate \cite{H,grangier}, but in principle they could be easily engineered  in microwave fields in coplanar transmission lines as it has been demonstrated recently \cite{martinis}. Previously \cite{omc} we have demonstrated an alternative proposal to generate a superposition
of  coherent states of a microwave field in a transmission line
resonator through the interaction with a single superconducting
charge qubit controlled by an external  single classical magnetic
pulse. By assuming a continuously varying magnetic pulse the generation can be almost deterministic. Since at low temperatures the
dissipative effects over the field are almost negligible during
the time interval the superposition is generated, one obvious question would be on the potentiality of this system for encoding quantum bits \cite{Mc1}.

 Assuming the circuit of Fig.  \ref{fig1}  we ask wether this strong regime can be optimally employed for a hybrid sort of quantum computation, with the qubits distributed in field qubits and solid states ones. Here we show a robust field qubit encoding in terms of a superposition of coherent state, allowing a hybrid implementation of quantum computing with the addition of solid state charge qubits. We show some applications of the encoding for both quantum computing and for generation of highly entangled states.
The paper is organized as follows. In Sec. II we describe the procedures to treat the time-dependent state dynamics. In Sec. III we review the procedure to generate a field superposition of coherent states and show how it can be employed for encoding a two-qubit quantum gate can be implemented.  In Sec IV we discuss how to generate deterministic single qubit rotations. In Sec. V we show how the procedures of Sec III can be extended to generate cluster states involving many solid state qubits. In Sec. VI we analyze the system relaxation due to external coupling to the environment. and finally in Sec VII a discussion concludes the paper.

\section{ Time-Dependent State Dynamics}

Let us consider the dynamics of the coherent state of radiation at the central transmission line (resonator) of the Fig. \ref{fig1}. The procedures were detailed in Ref. \cite{omc}, for a somewhat similar situation. We start our description by applying a rotation over the atomic basis  $\sigma_{z}\rightarrow\sigma_{x}$ and $\sigma_{x}\rightarrow-\sigma_{z}$ on Eq. (\ref{eq1}), so that we end up with the new basis ($\left|+\right\rangle$ and $\left|-\right\rangle$).  Moreover we apply a rotation $\exp\left[i\omega_{c}\left(a^{\dag}a+\sigma_{z}\right)t\right]$ over the atom and field Hamiltonian in order to describe it in a rotating frame as
\begin{eqnarray}
H^{\prime}&=& \hbar \Delta(t)\sigma_{z}+\hbar g\left(\tilde{\sigma}^{+}+\tilde{\sigma}^{-}\right)\left(\tilde{a}^{\dag}+\tilde{a}\right)
\label{88}
\end{eqnarray}
where $\tilde{a}^{\dag}=a^{\dag}\exp\left(i\omega_{c}t\right)$, $\tilde{a}=a \exp\left(-i\omega_{c}t\right)$, $\tilde{\sigma}^{+}=\sigma_{+}\exp\left(2i\omega_{c}t\right)$ and $\tilde{\sigma}^{-}=\sigma_{-}\exp\left(-2i\omega_{c}t\right)$. The detuning of the radiation field to the atomic transition is given by 
\begin{equation}
\Delta(t)\equiv \frac{E_{J}}{\hbar} \cos\left[\pi\frac{\Phi_{x}(t)}{\Phi_{0}}\right]- \omega_{c},\end{equation}
in terms of the time dependent  classical magnetic flux, $\Phi_{x}(t)$, externally applied to the SQUID of Fig. \ref{fig1}, the Josephson energy $E_J$ and the quantum of magnetic flux $\Phi_0=hc/2e$.  Since a situation of very large to large detuning is always assumed, we avoid the employment of a rotating wave approximation. We employ a  time-dependent perturbation theory, with $H_{0}=\hbar \Delta(t)\sigma_{z}$ and $V_{I}^{S}=\hbar g\left(\tilde{\sigma}^{+}+\tilde{\sigma}^{-}\right)\left(\tilde{a}^{\dag}+\tilde{a}\right)$, such that $H_{0}>>V_{I}^{S}$.
In our analysis consider that the field is always blue detuned from the atomic transition of frequency $\nu_a(t)=({E_{J}}/{\hbar}) \cos\left[\pi\frac{\Phi_{x}(t)}{\Phi_{0}}\right]$, as shown in Fig.    \ref{fig1}, where the frequency of atomic transition levels will vary, due to the time dependent magnetic pulse. With the extra condition of a number of less than one photon in the central transmission line, we can make approximations so that we can transfer information from the atom to the field with only a perturbation into the atomic system.

The formal solution
for the evolution operator reads as
\begin{eqnarray}
U(t,t_{0})&=&1+\sum^{\infty}_{n=1}\frac{1}{(i\hbar)^{n}}\int^{t}_{t_{0}}dt_{1}
\int^{t_{1}}_{t_{0}}dt_{2}\ldots\int^{t_{n-1}}_{t_{0}}dt_{n}\nonumber\\
&&\times\tilde{V}^{I}(t_{1})\tilde{V}^{I}(t_{2})\ldots\tilde{V}^{I}(t_{n}),
\label{10}
\end{eqnarray}
with $\tilde{V}^{I}(t)\equiv
U_{0}^{\dag}\left\{g\left(\tilde{\sigma}^{+}+\tilde{\sigma}^{-}\right)\left(\tilde{a}^{\dag}+
\tilde{a}\right)\right\}U_{0}$,
  and
$$ U_{0}=\exp\left[i\sigma_{z}\int_{t_{0}}^{t}dt^{'}\Delta(t^{'})\right].$$
For short times (See ref. \cite{omc}), and when the field in the central line resonator is prepared in a
coherent state $|\alpha\rangle$ with an average number of photons
smaller than one, the following approximation can be made  $ 1+ \theta_{\pm}(t) a^\dagger a\approx
e^{\left[\theta_{\pm}(t) a^{\dag}a\right]}$, where $\theta_{+}(t)$ and $\theta_{-}(t)$ are  the respective contributions in the dynamics of coherent state of radiation on the transmission line when the artificial atom is in the states $\left|+\right\rangle$ e $\left|-\right\rangle$. The action of the evolution operator over a initial state of the artificial atom and a coherent state of the field can be approximated as (preserving terms up to second order of perturbation)
\begin{eqnarray}
U(t)&&\left|-\right\rangle\left|\alpha\right\rangle\longrightarrow\left|-\right\rangle\left|\alpha \exp\left[\theta_{-}(t)\right]\right\rangle,\\
\label{3}
U(t)&&\left|+\right\rangle\left|\alpha\right\rangle\longrightarrow\left|+\right\rangle\left|\alpha \exp\left[\theta_{+}(t)\right]\right\rangle.
\label{4}
\end{eqnarray}
  For the present calculations we employ the following classical magnetic flux,
\begin{equation}
\Phi_{x}(t)=\frac{A\Phi_{0}}{2}\exp\left[i(\nu t+\varphi)\right],
\label{5}
\end{equation}
where $A=0,7$ is a strength parameter, and $\nu$ is made variable to better study the model. Due to the presence of the imaginary part of the pulse, the first order term on the expansion remains small up to the first semi-period, thus not affecting the dynamics. This contrasts to what was observed in \cite{omc} with a real oscillatory pulse.
Since the argument of $\Phi_{x}(t)(t)$ is kept small we approximate $\cos\left[\pi \frac{\Phi_{x}(t)}{\Phi_{0}}\right]\approx 1-\left(\pi\frac{\Phi_{x}(t)}{\Phi_{0}}\right)^{2}/2$. The appropriate regime for experimental implementation is $K_{B}T<<E_{J}<<E_{C}<<\delta$, so for typical  temperatures of $T\approx30mK$, $K_{B}T\approx3\mu eV$, $E_{J}/ \hbar \approx 15,9 \times 10^{10} Hz$, and the single electron charging energy $E_{C}=250\mu eV$. Here $\delta\approx458,3 \mu eV$ is \textit{gap} of the superconducting energy.  The typical frequency for the field is chosen as $\omega_{c}=70,7\times10^{10}Hz$.
%

\begin{figure}[!ht]
\begin{center}
\setlength\fboxsep{0.5pt}
\setlength\fboxrule{0.5pt}
\includegraphics[scale=0.98]{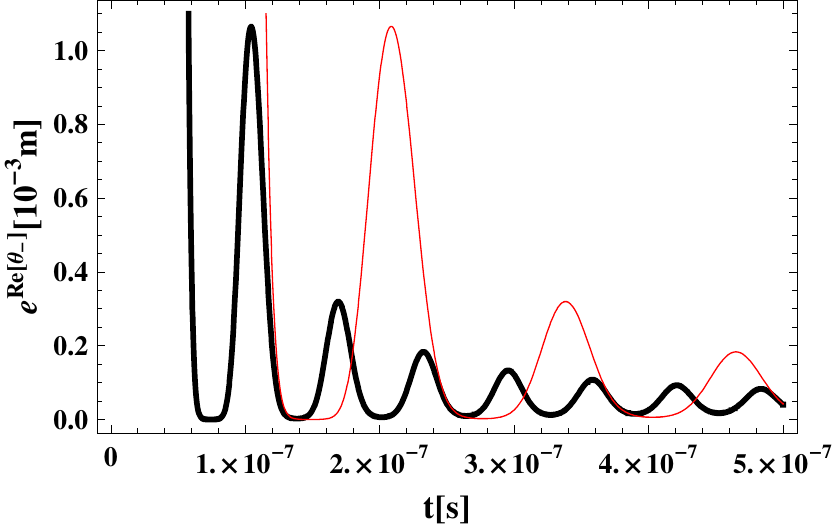}
\end{center}
\caption{\label{aecf} Variation of the amplitude of the coherent state field $e^{Re[\theta_{-}]}$ when the atomic state is $\left|-\right\rangle$ as a function of time for two values of the frequency of the classical magnetic flow $\nu$: red line for $\nu=8\times\pi\times10^{6}Hz$ and black line for $\nu=16\times\pi\times10^{6}Hz$. }
\end{figure}

\begin{figure}[!ht]
\begin{center}
\setlength\fboxsep{0.5pt}
\setlength\fboxrule{0.5pt}
\includegraphics[scale=0.2]{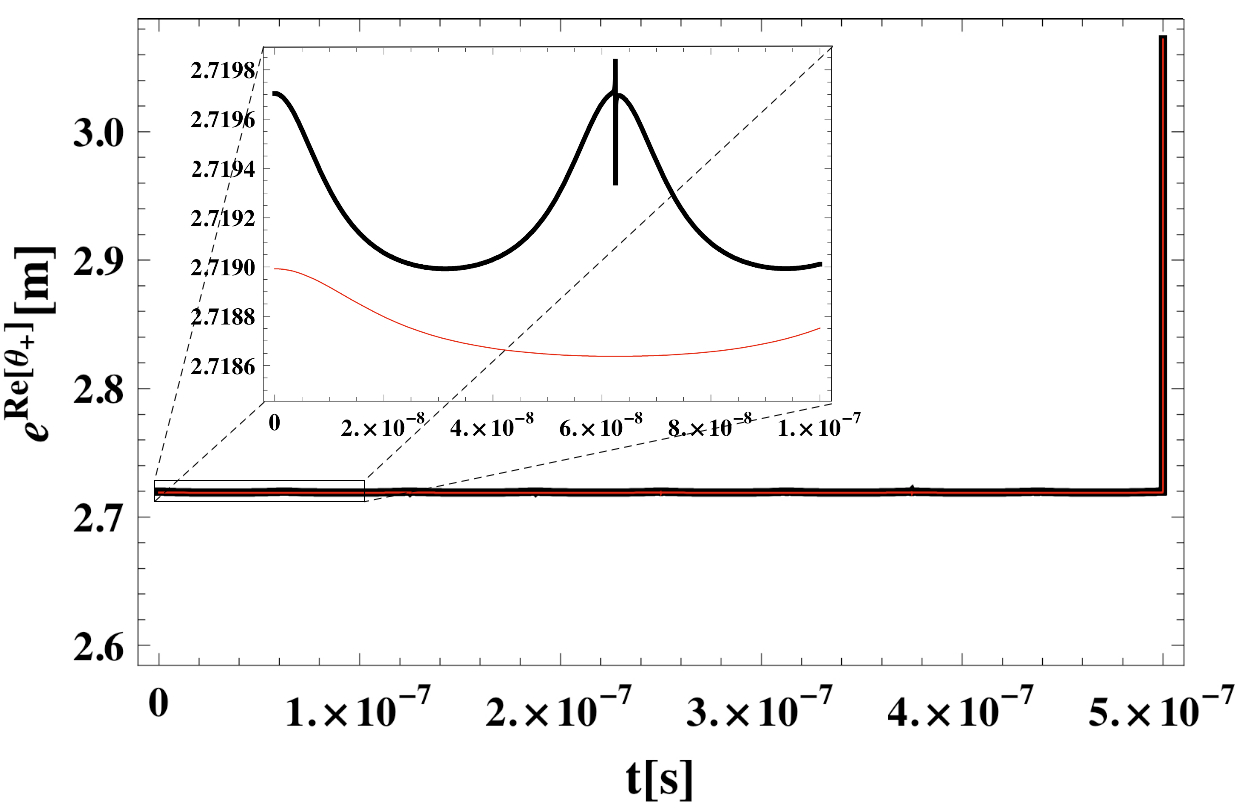}
\end{center}
\caption{\label{aece} Variation of the amplitude of the coherent state field $e^{Re[\theta_{+}]}$ when the atomic state is $\left|+\right\rangle$ as a function of time for two values of the frequency of the classical magnetic flow $\nu$: red line for $\nu=8\times\pi\times10^{6}Hz$ and black line for $\nu=16\times\pi\times10^{6}Hz$.}
\end{figure}

The real and imaginary part of the average $\left\langle \alpha,\pm\right|U(t=(2\nu)^{-1},n)\left|\alpha,\pm\right\rangle=\left\langle \alpha\right|U^{\pm}(t=(2\nu)^{-1},n)\left|\alpha\right\rangle$  allows to track the effectiveness of the approximation $e^{\left[\theta_{\pm}(t) a^{\dag}a\right]}$ for the situation  $1+ \theta_{\pm}(t) a^\dagger a$. For a short time (short here is to be considered until the first half period of oscillation of the classical magnetic field) and an average number of photons around $\bar{n}\approx0.4-0.5$, the approximation is invariably good. For shorter times than the half period of oscillation of the classical magnetic flux the average photon can be increased. Other  frequency $\nu$ values in $\Phi_{x}(t)$ can be employed within the validity of the approach, but do not change the properties mentioned, being the difference  only in the period of oscillation. This is exemplified in Figs.  (\ref{aecf}) and (\ref{aece}), which show the variation of $e^{Re[\theta_{-}]}$ and $e^{Re[\theta_{+}]}$, respectively,  as a function of time for two values of the classical magnetic flux  frequency $\nu$ applied over the artificial atom. The amplitude $e^{Re[\theta_{-}]}$ shows an oscillatory decay profile, with a frequency of oscillation lower than the corresponding $\nu$. At long times it stabilizes around a very small amplitude value, which increases with $\nu$, but still remains very small. This same feature is not observed for $e^{Re[\theta_{+}]}$. Although oscillating at the same rate of the previous  case, it does not decay to a constant value. However since we are always interested in a short time description, we can see that the amplitude is oscillating around a fixed value, which for practical purposes can be considered as constant.

\begin{figure}[!ht]
\begin{center}
\setlength\fboxsep{0.5pt}
\setlength\fboxrule{0.5pt}
\includegraphics[scale=0.98]{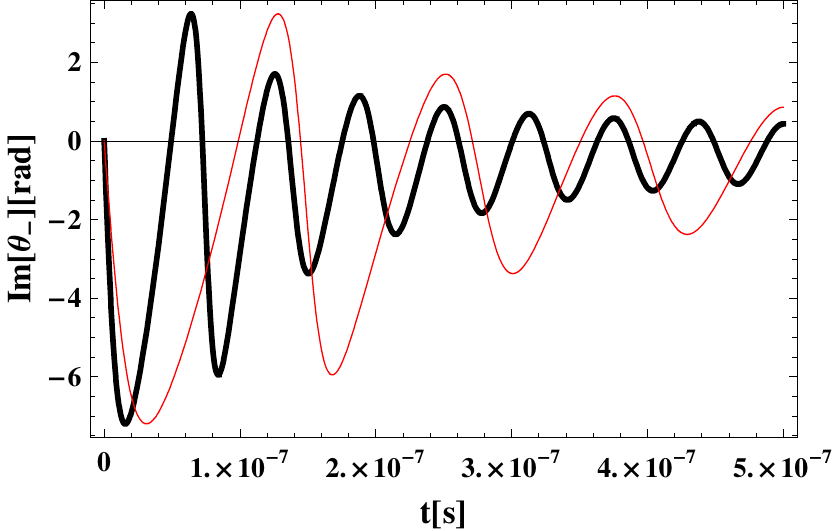}
\end{center}
\caption{\label{fecf}Change of the phase of the coherent state field $Im[\theta_{-}]$ when the atomic state is $\left|-\right\rangle$ as a function of time for two values of the frequency of the classical magnetic flow $\nu$: red line for $\nu=8\times\pi\times10^{6}Hz$ and black line for $\nu=16\times\pi\times10^{6}Hz$.}
\end{figure}
\begin{figure}[!ht]
\begin{center}
\setlength\fboxsep{0.5pt}
\setlength\fboxrule{0.5pt}
\includegraphics[scale=0.32]{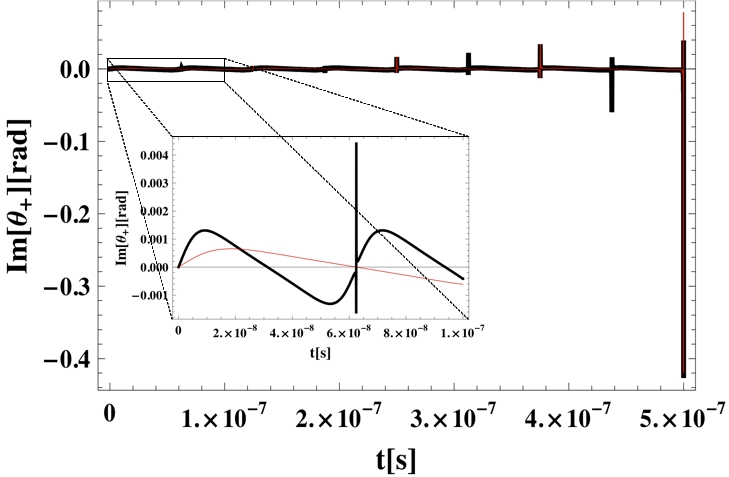}
\end{center}
\caption{\label{fece} Change of the phase of the coherent state field $Im[\theta_{+}]$ when the atomic state is $\left|+\right\rangle$ as a function of time for two values of the frequency of the classical magnetic flow $\nu$: red line for $\nu=8\times\pi\times10^{6}Hz$ and black line for $\nu=16\times\pi\times10^{6}Hz$.}
\end{figure}

Figs. (\ref{fecf}) and (\ref{fece}) show the variation of the phase $Im[\theta_{-}]$ and $Im[\theta_{+}]$, respectively, as a function of time for the same two values of the frequency $\nu$. The regime of $\nu=0$ is far from the  Stark A. C. \cite{JG} cumulative frequency due to large detuning between the frequency of the radiation field of the resonator and the frequency of transition of atomic energy levels. What is seen here for $\nu\neq0$ is that for each half period of the classical magnetic flux applied to the atom there is an accumulated phase value $\pi$ to the phase $Im[\theta_{-}]$. In the case of phase $Im[\theta_{+}]$, the same situation dos not apply. So we can say that the charge qubit and the field interaction imprint information to the amplitude of the state $\left|+\right\rangle$, and the phase for the state $\left|-\right\rangle$. 
Assembling these results, we can say that: (i) The oscillation frequency $\nu$ of the pulse $\Phi_{x}(t)$ is determinant for the time to shut down the pulse itself for achieving a required accumulated phase in one state $\left|-\right\rangle$. (ii) The phase accumulated in a single atomic state $\left|-\right\rangle$ is related to the difference in frequency of the quantized radiation transmission line frequency with the frequency of atomic transition. (iii) If the detuning of the frequency of the quantized radiation transmission line to the atomic transition frequency is very large (the strongly dispersive regime), the dynamics of the accumulated phase cannot assume a rotating wave approximation, and can be well described via a time-dependent perturbation theory. If the detuning of the frequency of the quantized radiation transmission line to the atomic transition frequency is very small (the strong resonant regime) the dynamics of the accumulated phase can be properly described with a rotating wave approximation, and is described like a A.C. Stark shift. (iv) The phase $\varphi$  in Eq. (\ref{5}) is to adjust the approach nullifying the first-order terms of the evolution operator at $U(t=1/2\nu)$. This results in a slight delay in the accumulation of the phase for $\left|-\right\rangle$.

\section{Encoding states and two-qubit logical operations}

 The artificial atom, composed by only two levels is an obvious charge qubit, where single qubit operations can be implemented through a coupling to a classical radiation field, when it is uncoupled from the central field resonator. We postpone this description to the next section, and consider now the encoding of field qubits and 2-qubits operations, since they are much simpler at this stage. We consider  $\{|0\rangle,|1\rangle\}=\{|g\rangle,|e\rangle\}$ as the appropriate computational basis for the solid state qubit.
 From now on we set in $\Phi_{x}(t)$,  Eq. (\ref{5}),  a frequency  $\nu=16 \pi\times10^{6}Hz$ during a time interval $\mathcal{T}=62,5ns$ (pulse time length), to imprint a conditional phase over the field coherent state, dependent on the state of the artificial atom. Other frequencies can be applied during different times resulting as well in the generation of other superposition states. Previously to the pulse we prepare the artificial atom in the ground state $\left|0\right\rangle$, here represented in  the $\{\left|+\right\rangle,\left|-\right\rangle\}$ basis and the field is considered in a coherent state
\begin{equation}
\left|\psi(0)\right\rangle=\left|0\right\rangle\otimes\left|\alpha\right\rangle=\frac{1}{\sqrt{2}}{\left(\left|-\right\rangle+\left|+\right\rangle\right)}\otimes\left|\alpha\right\rangle.
\label{400}
\end{equation}
With a pulse of $\mathcal{T}=62,5ns$ a $\pi$ phase is imprinted over the field coherent  state when the atom is in the $|-\rangle$ state so that \begin{eqnarray}
\left|\psi(\mathcal{T})\right\rangle&=&\frac{1}{\sqrt{2}}{\left(U(\mathcal{T})\left|-\right\rangle\otimes\left|\alpha\right\rangle+U(\mathcal{T})\left|+\right\rangle\otimes\left|\alpha\right\rangle\right)}\nonumber\\
&&\approx \frac{1}{\sqrt{2}}{\left(\left|-\right\rangle\otimes\left|-\alpha\right\rangle+\left|+\right\rangle\otimes\left|\alpha\right\rangle\right)},
\label{401}
\end{eqnarray}
or writing in the old basis $\left|0\right\rangle$ e $\left|1\right\rangle$,
\begin{eqnarray}
\left|\psi(\mathcal{T})\right\rangle&\approx&\frac{1}{2}\left[{\left|0\right\rangle\otimes\left(\left|\alpha\right\rangle+\left|-\alpha\right\rangle\right)}\right.\nonumber\\
&&\left.+{\left|1\right\rangle\otimes\left(\left|-\alpha\right\rangle-\left|\alpha\right\rangle\right)}\right],
\label{402}
\end{eqnarray}
so that if the atom is detected in $\left|0\right\rangle$ the field is left in the unnormalized state $\widetilde{|\alpha_+\rangle}=\left|\alpha\right\rangle+\left|-\alpha\right\rangle$, while if the atom is detected in $\left|1\right\rangle$ the field is left in the unnormalized state $\widetilde{|\alpha_-\rangle}=\left|-\alpha\right\rangle-\left|\alpha\right\rangle$. Normalizing with the appropriate probability of detection of $|0\rangle$ or $|1\rangle$, given respectively by $P_0(\mathcal{T})=Tr \left\{|0\rangle\langle 0|\psi(\mathcal{T})\rangle \langle\psi(\mathcal{T})|\right\}=(1+e^{-2|\alpha|^2})/2$, and  $P_1(\mathcal{T})=Tr \left\{|1\rangle\langle 1|\psi(\mathcal{T})\rangle \langle\psi(\mathcal{T})|\right\}=(1-e^{-2|\alpha|^2})/2$ we end up with the so-called even and odd coherent states, $|\alpha_+\rangle=(\left|\alpha\right\rangle+\left|-\alpha\right\rangle)/N_+$ and $|\alpha_-\rangle=(\left|-\alpha\right\rangle-\left|\alpha\right\rangle)/N_-$, respectively, with $N_{\pm}=\sqrt{2(1\pm e^{-2|\alpha|^2})}$.  The measurement on the artificial atom can be done by a single electron transistor \cite{eeq}, and detection is fully probabilistic in eigenstate $\left|0\right\rangle$ or $\left|1\right\rangle$. 

   The states $|\alpha_+\rangle$ and $|\alpha_-\rangle$ are orthogonal and can be though to encode a field qubit \cite{Mc1} as
\begin{eqnarray}
\left|0\right\rangle_{L}&&\equiv|\alpha_+\rangle,\\
\label{1000}
\left|1\right\rangle_{L}&&\equiv|\alpha_-\rangle.
\label{1001}
\end{eqnarray}
The advantage of this encoding is to be resilient to one photon losses, since the only possible noise in this case is  a symmetric bit flip. This kind of errors are less demanding to be corrected and one could think of these field qubits as the ones appropriate for information storage, while the solid state qubits are mostly relevant for rapid processing. It is easy to see that the normalized transformation \begin{equation}
|0\rangle|\alpha\rangle\rightarrow \frac{1}{\sqrt{2}}\left(|0\rangle|0\rangle_L+|1\rangle|1\rangle_L\right),\label{ewq}\end{equation}
really corresponds to a probabilistic Hadamard gate over the field qubit. By applying the pulse over the solid state qubit again the state (\ref{ewq}) returns to $|0\rangle|\alpha\rangle$, with similar transformation if we have started with the solid state qubit in state $|1\rangle$.
In a similar fashion when a detection on the solid state qubit population is followed by a second pulse a 2-qubit gate can be realized. Once the first classical magnetic pulse of magnetic flux is applied over the charge qubit, leading to Eq. (\ref{ewq}) we can detect its the state as logical $|0\rangle$ or $|1\rangle$ with equal probability, and so the field qubit $|0\rangle_L$ or $|1\rangle_L$ state is generated. After that generation, at a second stage we can prepare the charge qubit into another state $|0\rangle$ or $|1\rangle$ and after combining all the possible entries for both qubits we can apply a second pulse (similarly to what is described in the last section) to obtain
\begin{eqnarray*}
{\left|0\right\rangle\otimes\left|0\right\rangle_{L}}\rightarrow\frac{U(\mathcal{T})}{\sqrt{2}}\left(\left|-\right\rangle+\left|+\right\rangle\right)\otimes\left|0\right\rangle_{L}\rightarrow{\left|0\right\rangle\otimes\left|0\right\rangle_{L}}
\label{1002}\\
{\left|1\right\rangle\otimes\left|0\right\rangle_{L}}\rightarrow\frac{U(\mathcal{T})}{\sqrt{2}}\left(\left|-\right\rangle-\left|+\right\rangle\right)\otimes\left|0\right\rangle_{L}\rightarrow{\left|1\right\rangle\otimes\left|0\right\rangle_{L}}
\label{1003}\\
{\left|0\right\rangle\otimes\left|1\right\rangle_{L}}\rightarrow\frac{U(\mathcal{T})}{\sqrt{2}}\left(\left|-\right\rangle+\left|+\right\rangle\right)\otimes\left|1\right\rangle_{L}\rightarrow{\left|1\right\rangle\otimes\left|1\right\rangle_{L}}
\label{1004}\\
\underbrace{\left|1\right\rangle\otimes\left|1\right\rangle_{L}}_{input}
\rightarrow\frac{U(\mathcal{T})}{\sqrt{2}}\left(\left|-\right\rangle-\left|+\right\rangle\right)\otimes\left|1\right\rangle_{L}\rightarrow\underbrace{\left|0\right\rangle\otimes\left|1\right\rangle_{L}}_{output}
\label{1005}
\end{eqnarray*}
\begin{figure}[!ht]
\includegraphics[scale=0.35]{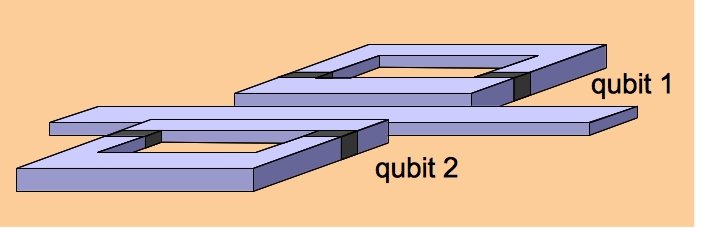}
\caption{\label{fig7} Capacitive coupling of two solid state qubits 1 and 2 to a single transmission line.}
\end{figure}
By comparing the input to the output we see that the action of the second pulse over the system as completely analogous to the CNOT gate, taking the field as a control qubit and the charge qubit as a target.

\begin{figure}[!ht]
\includegraphics[scale=0.3]{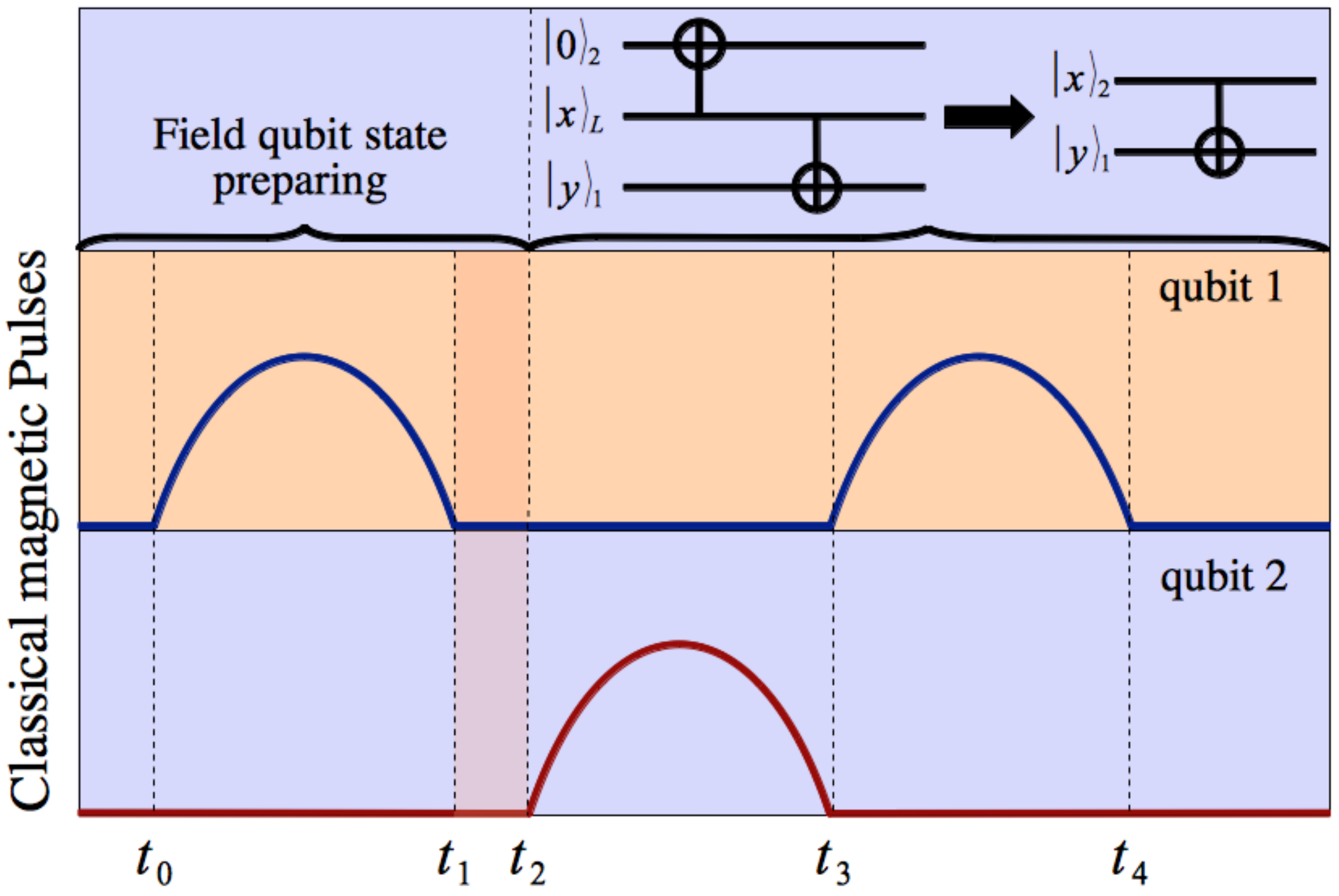}
\caption{\label{fig8} Sequence of classical magnetic pulses over solid state qubits 1 and 2, for implementation of a field-induced CNOT gate. Time duration of pulses are set to $\mathcal{T}=t_1-t_0=t_3-t_2=t_4-t_3$. The time interval $\Delta t_m=t_2-t_1$ is the duration of measurement over qubit 1 for generation of the field superposition encoding state.}
\end{figure}

Moreover, the CNOT between \textit{two distinct solid state qubits} can be implemented if previously (or after) to the CNOT operation, described above, the field is made to interact in the same way with a second solid state two-level artificial atom  (See Figs. \ref{fig7} and \ref{fig8}). For that the classical magnetic pulse has to be applied over this second solid state qubit only, in order to avoid any interaction with the first solid state qubit. It is straightforward to see that if this second qubit is initially prepared in the $|0\rangle_2$ state, after the interaction with the field qubit it will be transformed as
\begin{eqnarray}
|0\rangle_2|0\rangle_L &\longrightarrow&|0\rangle_2|0\rangle_L,\nonumber\\
|0\rangle_2|1\rangle_L& \longrightarrow&|1\rangle_2|1\rangle_L.
\end{eqnarray}
So now it is obvious that by looking only at the states of the first and the second solid state qubit, after the interaction of both with the field qubit, we en up with the CNOT truth table, where the second qubit acts as the control
\begin{eqnarray*}
{\left|0\right\rangle_1\otimes\left|0\right\rangle_{2}}\rightarrow{\left|0\right\rangle_1\otimes\left|0\right\rangle_{2}}
\label{1002}\\
{\left|1\right\rangle_1\otimes\left|0\right\rangle_{2}}\rightarrow{\left|1\right\rangle_1\otimes\left|0\right\rangle_{2}}
\label{1003}\\
{\left|0\right\rangle_1\otimes\left|1\right\rangle_{2}}\rightarrow{\left|1\right\rangle_1\otimes\left|1\right\rangle_{2}}
\label{1004}\\
{\left|1\right\rangle_1\otimes\left|1\right\rangle_{2}}
\rightarrow{\left|0\right\rangle_1\otimes\left|1\right\rangle_{2}}
\label{1005}
\end{eqnarray*}

\section{Two transmission line coupled to a Qubit as a quantum switch - one qubit operations}

One-qubit operations over the solid state qubit can be performed through their coupling to classical fields. It can only be performed in the present proposal with the addition of a second auxiliary classical field. That could be the radiation of a secondary transmission line, capacitively coupled to the qubit, or even a secondary classical mode of the same transmission line. In this section will show the interaction between a single solid state qubit and two radiation modes. This proposal follows closely the discussion in reference \cite{solano}. The Hamiltonian in the $\{|0\rangle,|1\rangle\}$ reads as follows
\begin{eqnarray*}
H_2&=&\hbar\omega_{a}a^{\dag}a+E_{J}cos\left(\pi\frac{\Phi_{x}(t)}{\Phi_0}\right)\sigma_{x}+\hbar g_{a}\sigma_{z}(a^{\dag}+a)\nonumber\\
&&+\hbar\omega_{b}b^{\dag}b+\hbar g_{b}\sigma_{z}(b^{\dag}+b)+\hbar g_{ab}(a^{\dag}+a)(b^{\dag}+b),
\end{eqnarray*}
where $g_{a}=\left(eC_{g}^{a}/\hbar(C_{J}+C_{g}^{a})\right)\times\sqrt{\hbar\omega_{a}/L_{a}c_{a}}$, $g_{b}=\left(eC_{g}^{b}/\hbar(C_{J}+C_{g}^{b})\right)\times\sqrt{\hbar\omega_{b}/L_{b}c_{b}}$, $g_{ab}=\left(e^{2}C_{g}^{a}C_{g}^{b}/\hbar(C_{J}+C_{g}^{a})\right)\times\sqrt{\hbar^{2}\omega_{a}\omega_{b}/L_{a}L_{b}c_{a}c_{b}}$. $C_{g}^{a}$ and $C_{g}^{b}$ are charges capacitances; $C_{J}$ is the Josephson capacitance; $\omega_{a}$ and $\omega_{b}$ are the frequencies of electromagnetic field radiation  modes, with respective $L_{a}$ and $L_{b}$ transmission line  lengths, being $c_{a}$ and $c_{b}$ the capacitance densities of the transmission lines.
For our case, the secondary transmission line (or secondary mode of the same transmission line) has a loss rate  larger than the pump rate so that the radiation can be considered as being classical, such as $b\rightarrow\left\langle b\right\rangle e^{-i\omega_{b}t}$ and $b^{\dag}\rightarrow\left\langle b\right\rangle e^{i\omega_{b}t}$, and the Hamiltonian can be written as
\begin{eqnarray}
H_2&=&\hbar\omega_{a}a^{\dag}a+\hbar\omega_{b}\left\langle b\right\rangle^{2}+E_{J}cos\left(\pi\frac{\Phi_{x}(t)}{\Phi_0}\right)\sigma_{x}\nonumber\\
&&+\hbar g_{a}\sigma_{z}(a^{\dag}+a)+2\hbar g_{b}\left\langle b\right\rangle cos(\omega_{b}t)\sigma_{z}\nonumber\\
&&+2\left\langle b\right\rangle\hbar g_{ab}cos(\omega_{b}t)(a^{\dag}+a).
\end{eqnarray}
In that fashion the secondary mode serves as a classical pump for the quantized radiation mode. For large transmission line lengths (in our case, the order of cm) and small load capacitances, the two modes interaction given by $g_{ab}$ can be neglected, and the relevant Hamiltonian reads\begin{eqnarray}
H_2&\approx&\hbar\omega_{a}a^{\dag}a+\hbar\omega_{b}\left\langle b\right\rangle^{2}+E_{J}cos\left(\pi\frac{\Phi_{x}(t)}{\Phi_0}\right)\sigma_{x}\nonumber\\
&&+\hbar g_{a}\sigma_{z}(a^{\dag}+a)+2\hbar g_{b}\left\langle b\right\rangle cos(\omega_{b}t)\sigma_{z},
\end{eqnarray}
and the solid state qubit can be rotated by varying the intensity of the classical mode $\langle b\rangle$, such that $g_a\ll g_b\langle b\rangle$. Since this intensity is controlled by an external pump, it is easy to switch this rotation on and off. One important possibility of the local rotation is to perform the  Hadamard quantum gate $\left|0\choose{1}\right\rangle\rightarrow(\left|0\right\rangle\pm\left|1\right\rangle)/\sqrt{2}$.
An advantage of this secondary mode is the possibility to indirectly implement one qubit rotations on the field mode $a$ as well. This is conditioned on the coupling of both fields as well on the atomic detection.

\section{Generation of $N+1$ qubits entangled states}

By extension, it is possible to generalize the CNOT operation for several solid state qubits, in order to generate a highly entangled state such as  the "cluster state"  \cite{clus1,clus}. This is achieved by capacitively coupling $N$ ``artificial atoms'' with a single mode of electromagnetic radiation through the Hamiltonian
\begin{eqnarray}
H_{n}&=&\hbar\omega a^{\dag}a+\sum_{j=1}^{N}E_{J}^{j}cos\left[\pi\frac{\Phi_{x}^{j}(t)}{\Phi_{0}}\right]\sigma_{x}\nonumber\\
&&+\hbar g\sum_{j=1}^{N}\sigma_{z}^{j}(a^{\dag}+a)\nonumber\\
&&+H_{qq}
\label{}
\end{eqnarray}
{where $H_{qq}=-\sum_{j=1}^{N-1}H_j$ is the interaction between qubit - qubit, with 
 \begin{eqnarray*}
H_j=\frac{4E_{J}^{j}E_{J}^{j+1}}{E_{L}}cos\left[\frac{\pi\Phi_{x}^{j}(t)}{\Phi_{0}}\right]cos\left[\frac{\pi\Phi_{x}^{j+1}(t)}{\Phi_{0}}\right]\sigma_{y}^{j}\sigma_{y}^{j+1},\end{eqnarray*} where $E_{L}=C_{J}\Phi_{0}^{2}/C_{qb}\pi^{2}L$ is the inductive energy between two qubits and $C_{qb}=C_{J}C_{q}/(C_{J}+C_{q})$. For $E_{L}>>E_{J}^{j}E_{J}^{j+1}$, $H_{qq}$ is far less significant than the other terms of the Hamiltonian and can be neglected. Thus the system of $N$ charge qubits follows the same dynamical process for the generation of superposition states for a single atom interacting with a transmission line mode. $\Phi_{x}^{j}(t)$ enables this simultaneous action over the $N$ qubits, each one being similarly driven by independent classical magnetic pulses as in equation (\ref{5}).
\begin{figure}[!ht]
\setlength\fboxsep{0.2pt}
\setlength\fboxrule{0.2pt}
{\includegraphics[scale=0.42]{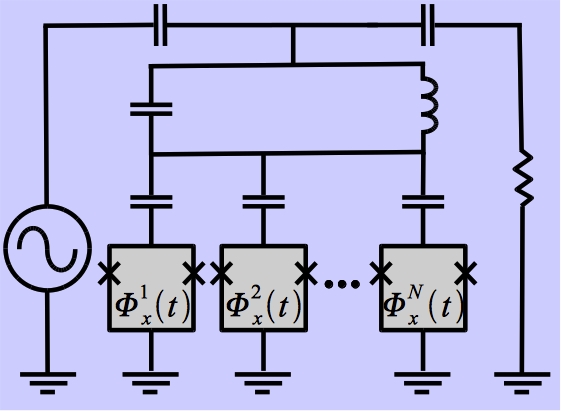}}
\caption{\label{fig9}Schematics for generation of $N$ entangled solid state qubits. The central transmission line (resonator) is capacitively coupled to the source and the drain of the quantized electromagnetic radiation, and also coupled capacitively to $N$ solid state qubit.}
\end{figure}
Assuming that all the magnetic pulses are simultaneous, for example, by starting from the initial state $\left|000\ldots0\right\rangle\left|\alpha\right\rangle$, soon after the application of the classical magnetics pulse $\Phi_{x}^{j}(t)$ on the $N$ ``artificial atoms'', during the same period $\mathcal{T}$, we will end up with a GHZ state of $N+1$ qubits
\begin{eqnarray}
\left|\psi_{N}(\mathcal{T})\right\rangle=\left|00\ldots0\right\rangle|0\rangle_L+(-1)^{N}\left|11\ldots1\right\rangle|1\rangle_L.
\label{}
\end{eqnarray}
By applying independent and non-simultaneous $\Phi_{x}^{j}(t)$ pulses, with distinct periods and initial qubit states one can obtain countless possible combinations of entangled states of the $N+1$ qubits.

\section{System relaxation}
In the current experiments with superconducting devices the decay time of the ``artificial atoms'' is much smaller than the field lifetime in the transmission line, so this last one can be neglected in a first instance. However the field coherence is also affected indirectly, due to the interaction with the artificial atom. The decay in the atom is due to the inductive character of the device, being characterized by an  ohmic bath, through the usual coupling of a bath of oscillators to the atom (system) \cite{spinb}. The study of this dissipative effect is well discussed in reference \cite{eeq} and its consequences for our specific problem in  \cite{omc}. For our main interest, after the relaxation time, all coherences will be destroyed and it extremely important that measurements be performed after the pulse operation and before the relaxation time of the system.\begin{figure}[!h]
\begin{center}
\setlength\fboxsep{0.5pt}
\setlength\fboxrule{0.5pt}
\subfloat[]{\label{p00}\includegraphics[scale=0.95]{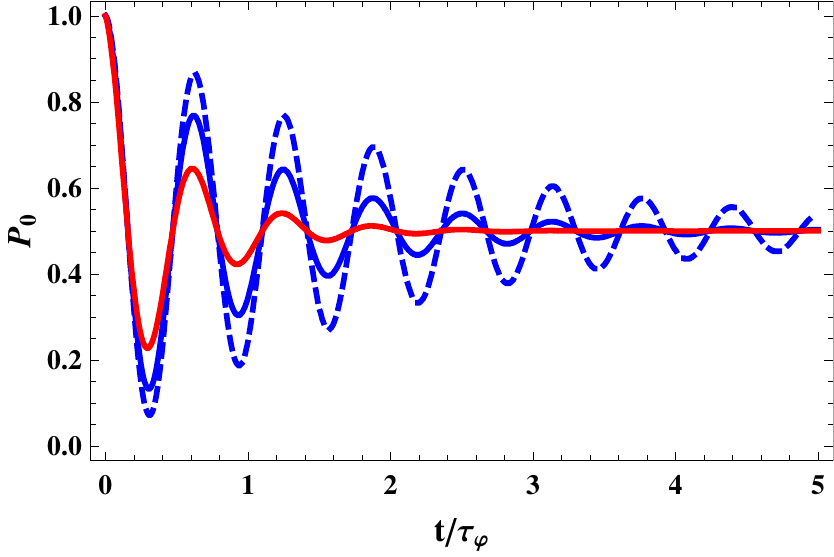}}\\
\subfloat[]{\label{p11}\includegraphics[scale=0.95]{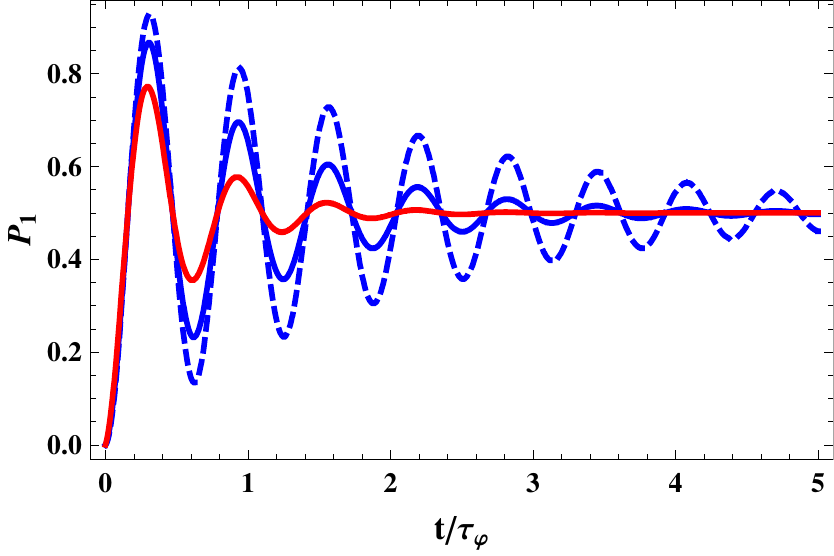}}\\
\subfloat[]{\label{pt}\includegraphics[scale=0.95]{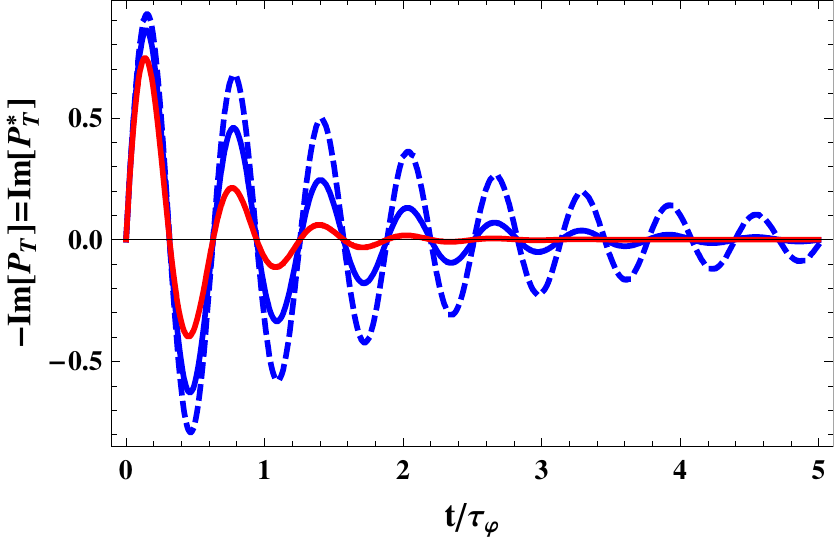}}\quad
\end{center}
\caption{(a) Probability of detection in the state, $\left|0\right\rangle$, $P_{0}(t)$ for the first atom and $P_{00}(t)$, to the second atom for different temperatures, the higher the temperature the higher the damping; (b) Probability of detection in the state $\left|1\right\rangle$, $P_{1}(t)$ for the first atom and $P_{11}(t)$, for different temperatures, the higher the temperature the higher the damping; (c) Imaginary part of the probability of measurement of the off-diagonal terms with different temperatures, the higher the temperature the higher the damping. The temperatures adopted were $T=10$ mK (dashed blue-line), $T=20$ mK (solid blue-line) and $T=40$ mK (solid red-line).}
\label{da}
\end{figure}
Following Ref. \cite{omc}, when the artificial atom two discrete levels coupled to an ohmic bath (allowing Markovian approximations on the solution of master equation), the probability to detect in in $|0\rangle$ or $|1\rangle$ state is given by
\begin{eqnarray}
P_{0\choose{1}}(t)&=&Tr\left\{\rho(t)\left|0\choose{1}\right\rangle\left\langle 0\choose{1}\right|\right\}\nonumber\\
&&=\frac12\left\{\tanh(\Lambda)
+[1-\tanh(\Lambda)] \exp(-t/\tau_{r})\right.\nonumber\\
&&\left.\pm\cos(2E_{J}t/\hbar) \exp(-t/\tau_{\varphi})\right\},\end{eqnarray}
while the atomic coherence is given by
\begin{eqnarray}
P_{T}(t)&=&Tr\left\{\rho(t)\sigma_{+}\right\}\nonumber\\
&&=-i\sin{(2E_{J}t/\hbar)}
\exp(-t/\tau_\varphi),\end{eqnarray} where $\Lambda\equiv2E/2k_{B}T$,
$\tau_{r}=[2\pi\beta E_{J} \coth(\Lambda)/\hbar]^{-1}$ is the
relaxation time and $\tau_{\varphi}=[\tau_{r}^{-1}/2+2\pi\beta
k_{B}T/\hbar]^{-1}$ is the dephasing time, with $\beta\approx0.001$  a
dimensionless parameter reflecting the strength of dissipation
\cite{eeq}, and $k_BT$ is the thermal energy, as usual.  For our purposes $E_{J}\gg k_B T$,  $\tanh(\Lambda)\approx
1$ and so \begin{eqnarray}
P_{0\choose{1}}(t)&=&\frac12[1\pm\cos(2E_{J}t/\hbar) \exp(-t/\tau_{\varphi})],\end{eqnarray}  reflecting the
probability to detect the qubit in the state $|0\rangle$ or
$|1\rangle$, respectively, at an instant $t$ after the classical
magnetic pulse. It is obvious that there is a high probability to find the atom in the state $|0\rangle$ if the measurement is performed shortly after the pulse (See Fig. \ref{da}).  Both probabilities are very sensitive to the increase of temperature, but more importantly is that for longer times both probabilities go to $1/2$. This means that for the case of generation of the encoding field states, there is an equiprobable chance to end with $|0\rangle_L$ or $|1\rangle_L$. Also for comparison we show in Fig. \ref{da}c the transition probability.
\begin{figure}[!h]
\setlength\fboxsep{0.5pt}
\setlength\fboxrule{0.5pt}
\subfloat[]{\label{fp00}\includegraphics[scale=0.95]{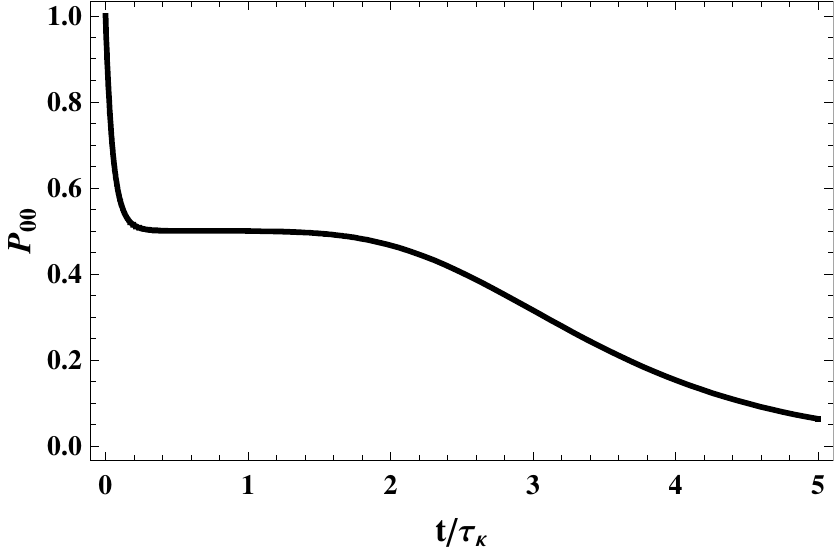}}\\
\subfloat[]{\label{fp10}\includegraphics[scale=0.95]{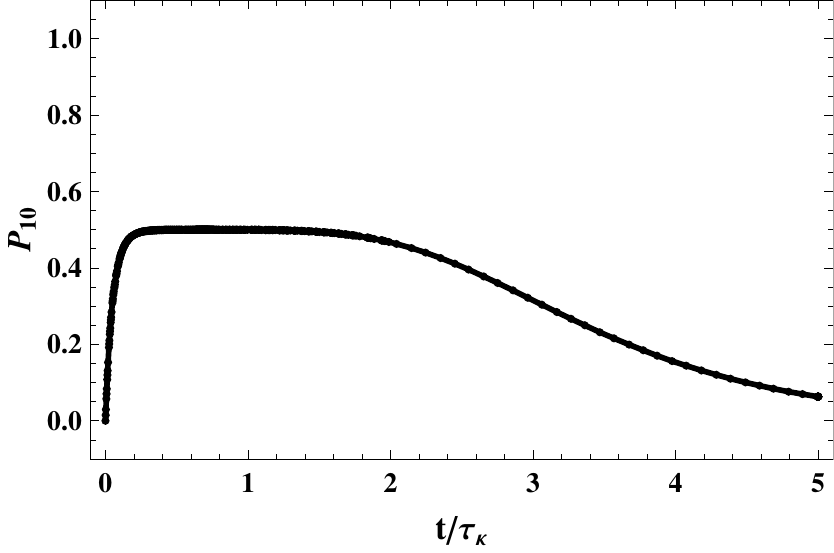}}
\caption{(a) Probability of measuring the atomic state in  $\left|0\right\rangle$ after the second magnetic pulse; (b) Probability of measuring the atomic state in  $\left|1\right\rangle$ after the second magnetic pulse, for $\alpha^{2}=10$.}
\label{df}
\end{figure}

In an optimistic experimental situation, where the decay time for the solid qubit would be larger than the field qubit, the dissipative effect over the coherence of the field can be probed by population measurements on the state of the artificial atom. Assuming that the effects of the environment over the field qubit is given by the usual amplitude damping channel, emulating the effect of a thermal reservoir at zero temperature \cite{Mc1,Mc2}. The coherence of the field is probed by considering that a second pulse of magnetic flux $\Phi_{x}(t)$ is applied over the device after a time interval $t$ \cite{H} . Then, we measure the probability of the solid state qubit to be in the state $\left|0\right\rangle$ or in the state $\left|1\right\rangle$. So if a second pulse is applied over the solid state qubit after the first measurement has detected $|0\rangle$, then we obtain the probability $P_{00}(t)$ for a subsequent detection on state $|0\rangle$ and $P_{10}(t)$ for detection on state $|1\rangle$. If is is initially measured in state $\left|0\right\rangle$, for $t/\tau_{\kappa}\leq1$ they are given by 
\begin{eqnarray*}
P_{j0}(t)&=&\frac{1}{2}\left[1+(-1)^j\frac{e^{-2\left|\alpha\right|^{2}e^{-\frac{t}{\tau_{\kappa}}}}+e^{-2\left|\alpha\right|^{2}\left(1-e^{-\frac{t}{\tau_{\kappa}}}\right)}}{1+e^{-2\left|\alpha\right|^{2}}}\right],
\label{d1}
\end{eqnarray*}
for $j=0,1$, and for  $t/\tau_{\kappa}\geq1$ both are given by
\begin{eqnarray*}
P_{j0}(t)&=&\frac{1}{2}\left[1-\frac{e^{-2\left|\alpha\right|^{2}e^{-\frac{t}{\tau_{\kappa}}}}+e^{-2\left|\alpha\right|^{2}\left(1-e^{-\frac{t}{\tau_{\kappa}}}\right)}}{1+e^{-2\left|\alpha\right|^{2}}}\right].
\label{d1}
\end{eqnarray*} 
Fig.  \ref{df} shows the two probabilities for sequential detection, which is reflected in the field qubit encoding  states $|0\rangle_L$ and $|1\rangle_L$, depending on the time interval between the first detection and the second magnetic pulse. Since this second pulse is the one being responsible for the CNOT operation, we see that the success of the whole operation is very sensitive to the decoherence effect.

\section{Conclusion}

In conclusion we have derived a qubit encoding in terms of a superposition of coherent states, allowing a hybrid implementation of quantum computing involving a transmission line radiation fields qubit and solid state charge qubits. We presented some applications of the encoding for both quantum computing and for generation of highly entangled states. For that we employed a time dependent classical magnetic pulse controlling the frequency of transition between two states of the solid state qubit.  A specific approach was developed in which information can be transferred from the superconducting device to the field and back to the device, creating superposition states of the coherent state of the resonator field.  For that some reasonable approximations were made: $(i)$ the average number of photons in the encoding field must be small, $(ii)$ a range of very short time of operation must be implemented, for both the generation of the field encoded qubit as well as for the computational operations. This is performed with a single perturbation on the qubit (a single pulse of magnetic flux). Analyzing the dissipation, the solid state qubits are more sensitive to decoherence, being the decoherence time related to  the inverse of the temperature of the thermal reservoir. On the other hand, in an optimistic situation, if we consider only the dissipation of the resonator, it possible to infer on the coherence time for the filed qubit via a sequential two pulses application followed by the solid state qubit population measurement. For the present analysis we employed worst experimental conditions, showing that our proposal could be implemented with present day technology.

\acknowledgements{We thank FAPESP and CNPq for financial support through the National Institute for Science and Technology for Quantum Information. We thank Amir Caldeira and Kyoko Furuya for several relevant discussions in developing this work. MCO acknowledges support by iCORE.


\begin{thebibliography}{99}

\bibitem{Milburn} D.F. Walls and Gerard J. Milburn, \textit{Quantum Optics}, (\textit{2ºed.}, Ed. Springer, 2007).
\bibitem{Scully} Marlan O. Scully and M. Suhail Zubairy, \textit{Quantum optics}, (Cambridge University Press, 1997).
\bibitem{Yamamoto} Y. Yamamoto, A. Imamoglu. \textit{Mesoscopic Quantum Optics}, (Wiley, 1999).


\bibitem{H} L. Davidovich, M. Brune, J. M. Raimond, and S. Haroche, Mesoscopic quantum coherences in cavity QED: Preparation and decoherence monitoring schemes, Physical Review A \textbf{53}, 1295 (1996); M. Brune, E. Hagley, J. Dreyer, X. Ma\^\i tre, A. Maali, C. Wunderlich, J. M. Raimond, and S. Haroche, Observing the Progressive Decoherence of the Meter in a Quantum Measurement, Physical Review Letters \textbf{77}, 4887 (1996) .
\bibitem{ref2} S. Haroche and J.M. Raimond, \textit{Exploring the Quantum: Atoms, Cavities, and Photons},  (Oxford University Press, NY,USA, 2006).




\bibitem{mae} A. Blais, Ren-Shou Huang, Andreas Wallraff, S. M. Girvin, and R. J. Schoelkopf, Cavity quantum electrodynamics for superconducting electrical circuits: An architecture for quantum computation, Physical Review A \textbf{69}, 062320 (2004).
\bibitem{mae1} A. Wallraff, D. I. Schuster, A. Blais, L. Frunzio, R.-S. Huang, J. Majer, S. Kumar, S. M. Girvin and R. J. Schoelkopf, Strong coupling of a single photon to a superconducting qubit using circuit quantum electrodynamics, Nature \textbf{431}, 162 (2004).
\bibitem{eeq} Y. Makhlin, G. Schon and A. Shnirman, Quantum-state engineering with Josephson-junction devices, Review  Modern Physics \textbf{73}, 357 (2001).
\bibitem{qubit}A. Blais, J. Gambetta, A. Wallraff, D.I. Schuster, S.M. Girvin, M.H. Devoret, and R.J. Schoelkopf, Quantum-information processing with circuit quantum electrodynamics, Physical Review A \textbf{75}, 032329 (2007).
\bibitem{solano} Matteo Mariantoni, Frank Deppe, A. Marx, R. Gross, F. K. Wilhelm, and E. Solano, Two-resonator circuit quantum electrodynamics: A superconducting quantum switch, Physical Review B, \textbf{78}, 104508 (2008).
\bibitem{solano1} G. Haack, F. Helmer, M. Mariantoni, F. Marquardt, E. Solano, Resonant quantum gates in circuit quantum electrodynamics, Physical Review B. \textbf{82}, 024514 (2010).
\bibitem{korotkov} A. N. Korotkov, Quantum Information Processing \textbf{8}, 51 (2009).
\bibitem{pashkin} Yu. A. Pashkin, O. Astafiev, T. Yamamoto, Y. Nakamura, and J. S. Tsai, Josephson charge qubits: a brief review, Quantum Information Processing, \textbf{8} 55 (2009).


























\bibitem{omc} O. P. de S\'{a} Neto, M. C. de Oliveira, A. O. Caldeira, Generation of superposition states and charge-qubit relaxation probing in a
circuit, Journal of Physics B: Atomic, Molecular and Optical Physics \textbf{44}, 135503 (2011).


\bibitem{JG} D. I. Schuster, A. A. Houck, J. A. Schreier, A. Wallraff1, J. M. Gambetta, A. Blais, L. Frunzio, J. Majer, B. Johnson, M. H. Devoret, S. M. Girvin and R. J. Schoelkopf, Resolving photon number states in a superconducting circuit, Nature \textbf{445}, 515 (2007).
\bibitem{entq}     L. DiCarlo,    M. D. Reed,    L. Sun,    B. R. Johnson,    J. M. Chow,    J. M. Gambetta,    L. Frunzio,    S. M. Girvin,    M. H. Devoret and R. J. Schoelkopf, Preparation and measurement of three-qubit entanglement in a superconducting circuit, Nature \textbf{467}, 574 (2010).

\bibitem{analog} N. Bergeal, R. Vijay, V. E. Manucharyan, I. Siddiqi, R. J. Schoelkopf, S. M. Girvin and M. H. Devoret, Analog information processing at the quantum limit with a Josephson ring modulator, Nature Physics \textbf{6}, 296 (2010).
\bibitem{Fock} H. Wang, M. Hofheinz, M. Ansmann, R. C. Bialczak, E. Lucero, M. Neeley, A. D. O. Connell, D. Sank, J. Wenner, A. N. Cleland e John M. Martinis. {Measurement of the Decay of Fock States in a Superconducting Quantum Circuit}, Physical Reveview Letters \textbf{101}, 240401 (2008).\bibitem{Fock2} Max Hofheinz, E. M. Weig, M. Ansmann, Radoslaw C. Bialczak, Erik Lucero, M. Neeley, A. D. O. Connell, H. Wang, John M. Martinis e A. N. Cleland, Generation of Fock states in a superconducting quantum circuit, Nature \textbf{454}, 310 (2008).
\bibitem{Mc1} M.C. de Oliveira and W.J. Munro, Quantum computation with mesoscopic superposition states, Physical Review A \textbf{61}, 042309 (2000).


\bibitem{grangier} A. Ourjoumtsev, R. Tualle-Brouri, J. Laurat, and P. Grangier, Generating Optical Schrödinger Kittens for Quantum Information Processing, Science \textbf{312}, 83 (2006).
\bibitem{martinis} Max Hofheinz, H. Wang, M. Ansmann, Radoslaw C. Bialczak, Erik Lucero, M. Neeley, A. D. O'Connell, D. Sank, J. Wenner, John M. Martinis, A. N. Cleland, Synthesizing arbitrary quantum states in a superconducting resonator, Nature \textbf{459}, 546 (2009).


\bibitem{clus1}R. Raussendorf and H. J. Briegel, A One-Way Quantum Computer, Physical Review Letters \textbf{86}, 5188 (2001). 

\bibitem{clus} M. A. Nielsen, Cluster-state quantum computation, Reports on Mathematical Physics \textbf{57}, 147 (2006).












\bibitem{spinb} A J Leggett, S. Chakravarty, A. T. Dorsey, Matthew P. A. Fisher, Anupam Garg and W. Zwerger, Dynamics of the dissipative two-state system, Review Modern Physics \textbf{59}, 1 (1987).
\bibitem{Mc2} M. C. de Oliveira, M. H. Y. Moussa e S. S. Mizrahi, Continuous pumping and control of a mesoscopic superposition state in a lossy QED cavity, Physical Review A \textbf{61}, 063809 (2000).

\end{thebibliography}
\end{document}